\documentclass[conference]{IEEEtran}
\newcommand{\ignore}[1]{}
\usepackage{booktabs}
\usepackage{pgfplots}
\usepackage{listings}
\usepackage{etoolbox}
\usepackage{parcolumns}
\usepackage{algorithm}
\usepackage[noend]{algpseudocode}
\usepackage{amsmath}
\usepackage{booktabs}
\usepackage{color}
\usepackage[export]{adjustbox}
\usepackage[utf8]{inputenc}
\newcommand{\tma}[1]{\textbf{\color{magenta}[tma: #1]}}

\ifCLASSINFOpdf
\else
\fi

\begin{document}
\title{{{\O}}zone: Efficient Execution with Zero Timing Leakage for Modern Microarchitectures}


\author{\IEEEauthorblockN{Zelalem Birhanu Aweke and Todd Austin}
\IEEEauthorblockA{University of Michigan}
\IEEEauthorblockA{\textit{Email: \{zaweke,austin\}@umich.edu}}\\[-4.5ex]}


\maketitle

\begin{abstract}
Time variation during program execution can leak sensitive information. Time variations due to program control flow and hardware resource contention have been used to steal encryption keys in cipher implementations such as AES and RSA. A number of approaches to mitigate timing-based side-channel attacks have been proposed including cache partitioning, control-flow obfuscation and injecting timing noise into the outputs of code. While these techniques make timing-based side-channel attacks more difficult, they do not eliminate the risks. Prior techniques are either too specific or too expensive, and all leave remnants of the original timing side channel for later attackers to attempt to exploit.  

In this work, we show that the state-of-the-art techniques in timing side-channel protection, which limit timing leakage but do not eliminate it, still have significant vulnerabilities to timing-based side-channel attacks. To provide a means for total protection from timing-based side-channel attacks, we develop Ozone, the first zero timing leakage execution resource for a modern microarchitecture. Code in Ozone execute under a special hardware thread that gains exclusive access to a single core’s resources for a fixed (and limited) number of cycles during which it cannot be interrupted. Memory access under Ozone thread execution is limited to a fixed size uncached scratchpad memory, and all Ozone threads begin execution with a known fixed microarchitectural state. We evaluate Ozone using a number of security sensitive kernels that have previously been targets of timing side-channel attacks, and show that Ozone eliminates timing leakage with minimal performance overhead.
\end{abstract}


% \fi
\IEEEpeerreviewmaketitle

\section{Introduction}

\ignore{Traditionally, security concerns have revolved around the software components of a system, since most attacks were concentrated on these components. In recent years however, more attention has been placed on hardware security vulnerabilities, both by the design community and the attacker community. This increased attention is the result of recent hardware security vulnerabilities, such as the Rowhammer bug \cite{Kim} or SSD read-after-delete attacks \cite{Wei}.

\subsection{Timing-Based Side-Channel Attacks}}

Perhaps the most fruitful hardware security vulnerability has been timing-based side-channel attacks. \ignore{A side-channel attack is one in which a non-secure observable property of a system reveals secret information inside.} A {\em timing-based side-channel attack}\footnote{Some taxonomies of side-channel attacks create distinct categories for cache and control flow attacks. Since these attacks are perpetrated to correlate secrets to the timing of the cache or control flow, we broadly address all of these attacks under the class of timing attacks.} is one in which the time it takes to perform specific operations reveals information about secrets within the system.
There are three characteristics of modern systems that enable timing-based side-channel attacks:

\noindent\textbf{1. Variable-latency operations:} In an effort to improve the performance of a system, common operations are made to take less time to execute than uncommon ones; thus, it becomes possible to infer the constituency of instructions a program executes simply by examining its latency. Furthermore, control flow, speculation, caches, and a variety of other system features render execution latencies that are a function of the program's code and data, thereby creating opportunities to learn about such things. Kocher used this characteristic to attack RSA authentication~\cite{Kocher}, after noting that early RSA implentations performed different amounts of work for "0" and "1" key bits.

\noindent\textbf{2. Resource sharing:} To keep costs in check, functional and storage resources in a modern microarchitecture are shared concurrently among threads and processes; thus, it becomes possible for an attacker program to create contention on these resources and examine, by virtue of its own performance, the extent to which the shared resources are utilized by the victim program. \ignore{An example of a resource sharing attack is the Flush+Reload attack \cite{Yarom}, which uses shared code in dynamic libraries to enable an attack on OpenSSL's RSA implementation. By forcing shared library code out of the cache, the attacker can view which code segments were executed (and when) for each RSA authentication request. This attack is so powerful it can extract the RSA private key in only a few authentication requests.} Examples of resource sharing timing side-channel attacks include L1 instruction cache based attacks~\cite{Aciicmez_rsa}, L1 data cache based attacks~\cite{Osvik}\ignore{,Zhang}, branch predictor based attacks~\cite{BranchP} and most recently last-level cache based attacks~\cite{LLC,Yarom}.

\noindent\textbf{3. Fine-grained performance monitoring:} To facilitate debugging and code optimization, modern systems provide high-precision timing facilities, {\em e.g.}, Intel's {\em rdtsc} user-level cycle counter read operation, and performance monitoring capabilities. \ignore{\cite{Intel_sdm}, {\em e.g.}, ARM's PMU registers which can monitor cache and branch predictor performance \cite{ARM_sdm}.} While most timing side-channel attacks make use of high-precision timing facilities, performance counters ({\em e.g.}, cache miss counters) have also been used in the past ~\cite{Uhsadel}.\ignore{, such as Irazoqui's S\$A attack on AES that tracked last-level cache miss rates \cite{Irazoqui}.}
\ignore{to Irazoqui's S\$A attack on AES \cite{Irazoqui}   utilizes cache performance counters to perpetrate an attack via the last-level cache on AES running even on a different core (with a shared last-level cache) and a different virtual machine.}
%\end{enumerate}

\ignore{While the related work section of this paper details mitigation techniques in detail,} 

\subsection{Mitigating Timing Attacks}
There is a large body of work that has proposed various techniques to mitigate timing-based side-channel attacks. We broadly note that these techniques work to address each of the above three system characteristics that leak timing information. Variable-latency operations can be mitigated through a variety of software and/or hardware measures to reduce the data-dependent latency of a program. For example, some efforts to reduce control flow dependence on data utilize if-conversion and predicated execution to eliminate control hammocks\footnote{A control hammock is a CFG construct where control diverges from a single point based on a predicate and then reconverges again to a single point.  This construct is typically formed by IF and SWITCH statements.} \cite{Coppens,Molnar,Rane}. Other efforts work to lock down critical data into the cache \cite{Wang} or a scratchpad memory \cite{Liu} so as to reduce memory latency variability.  While these techniques work to limit the degree of data-dependent timing variation in programs, they do not completely eliminate it since a wide range of microachitectural features continue to perform data-dependent optimizations, leading to attacks via branch predictors \cite{BranchP} and cache conflicts \cite{Bernstein,Bonneau}, among others.

To address resource sharing, techniques have proposed to isolate programs to their own dedicated CPU and caches, so as to eliminate potential competition for resources from other aggressor threads, {\em e.g.} \cite{Braun}. While these techniques can eliminate aggressors at great cost, they still ultimately suffer from timing leakage due to variable-latency operations that remain on the single-threaded microarchitecture. As an example, Bernstein's cache attack on AES \cite{Bernstein} works by monitoring how AES-internal data cache conflicts vary with plaintext, thus making the aggressor in this case the AES code itself.

Finally, to mitigate the problems of fine-grained performance analysis, efforts have proposed time padding and obfuscation to obscure the true activities of a program, {\em e.g.}, \cite{Braun,Rane}. The former technique delays results from a program such that all results return after a fixed maximum delay. While this approach eliminates timing differences, it doesn't eliminate resource contention based attacks.\ignore{For example, Irazoqui's S\$A attack on AES would continue to work with time padding, since the timing analysis in this case is performed on the attackers code.}  Obfuscation techniques work by injecting immense amounts of noise into the execution of the program, through the use of extraneous memory accesses and path executions, {\em e.g.}, \cite{Rane}. \ignore{and oblivious RAM \cite{Liu}} While these techniques do attenuate timing leakage, they do so with enormous slowdown.

\subsection{Eliminating Timing Attacks with Ozone}

In this work, {\em we build the first execution resource, called Ozone, that exhibits {\bf\em zero timing leakage} on a modern microarchitecture}. Our approach is low cost, and it draws upon earlier work in quelling control and data timing leakage, but also introduces new techniques to silence microarchitectural timing leakage. In the Ozone execution environment, vulnerable codes, such as an AES kernel, will execute with a fixed (and well known) latency regardless of inputs. To achieve zero timing leakage, Ozone places restrictions on the code it can execute ({\em e.g.}, no CFG hammocks), thus, Ozone targets small applications and/or security-sensitive potions of applications.

Specifically, Ozone achieves zero timing leakage using the following measures to eliminate each of the three properties that enable timing attacks: {\em i)} the Ozone compiler guarantees that the instruction trace of an Ozone thread has a fixed number of instructions executing on a fixed control path, making the thread's execution always independent of input data, {\em ii)} Ozone codes execute under a special hardware thread that gains exclusive access to a single core's resources for a fixed (and limited) number of cycles during which it cannot be interrupted, {\em iii)} memory access is limited to a fixed size uncached scratchpad memory, and {\em iv)} all Ozone threads begin execution with a known fixed microarchitectural state.

\ignore{Given these four aspects of the Ozone execution framework, Ozone threads {\em i)} execute with fixed latency regardless of inputs on even complex microarchitectures, {\em ii)} are not subject to resource contention by other threads, and {\em iii)} reveal nothing by inspecting their execution with performance monitors, since their cycle-level execution is well known, deterministic, and unchanging. Consequently, Ozone threads have zero timing leakage, and thus, they are not subject to timing attacks.}

The ultimate challenge for Ozone is to demonstrate that its restrictive execution environment is capable of running codes that are vulnerable to timing-based side-channel attacks. \ignore{To implement zero timing leakage, a number of constraints are placed on how the programmer can express their algorithms, {\em e.g.}, fixed loop counts and no control hammocks.} To this end, we show in this work that we can efficiently execute a wide range of known-vulnerable codes inside the Ozone execution environment. {\em To our knowledge, our benchmarks represent the largest collection of known side-channel vulnerable codes analyzed to date, by a wide margin.} Indeed, the breadth of the codes analyzed makes a strong case for the Ozone approach to stopping timing-based side-channel attacks.

Specifically, we make the following novel contributions:
\begin{itemize}
\item We begin by demonstrating the significant value of zero timing leakage over state-of-the-art mitigation techniques that only attenuate timing leakage. We show, using the decade-old Bernstein's cache attack, that state-of-the-art timing side-channel mitigations are half measures. \ignore{An AES kernel with advanced protections for control and data timing leakage readily gives up the AES key to the Bernstein attack. In contract, the Ozone execution environment fully protects the AES key from a Bernstein cache attack.}

\item We detail the first {\em low-cost implementation} of an execution mode for modern microarchitectures with {\em zero timing leakage}, via the Ozone execution environment.  In addition, we demonstrate the utility of our approach by porting a broad array of side-channel vulnerable application components into the Ozone execution environment.

\item We analyze the security and efficiency of the Ozone zero timing leakage execution capability in the context of a modern microarchitecture, by examining the performance of the Ozone execution capability on the Gem5 detailed microarchitecture simulator. \ignore{We show that a wide array of timing-attack vulnerable benchmarks have zero timing leakage across a range of inputs. Additionally, we demonstrate that zero timing leakage is maintained in the context of exceptions, dynamic scheduling, interrupts, program bugs, and virtualization of the Ozone execution resource.}
\end{itemize}

The remainder of this paper is organized as follows. In Section 2, we demonstrate why state-of-the-art approaches to attenuate timing leakage are not sufficient to protect program secrets. Section 3 discusses the requirements for zero timing leakage execution and details the Ozone architectural enhancements. Section 4 gives details about the implementation of Ozone, and performs security analysis of the Ozone execution resource. Section 5 examines related work, and finally we conclude in Section 6.

\section{ Why Attenuating Timing Leakage is Not Enough}
%\section{Fatality of Small Leakages}
A number of techniques have been proposed to mitigate timing-based side-channel attacks. One popular mitigation technique is isolating security sensitive code execution. This is done by allocating private resources to security sensitive applications~\cite{Braun,cat,Wang}. In~\cite{cat,Wang} hardware and software mechanisms are used to partition caches into private regions to mitigate cache-based timing side-channel attacks. \ignore{These private regions are allocated to different processes, and data in private region of one process can not be evicted by concurrently running processes.} Another recent work by Braun {\em et. al.} \cite{Braun} targets both control-flow and resource contention based side-channel attacks. The method uses time padding through delay loops to account for time variations due to different program paths. To provide protection from cache-based timing attacks, the method reserves a core for the duration of execution of a security sensitive function, and it utilizes page coloring to reserve L3 cache resources. 

While these techniques provide significant attenuation of timing leakage, they all still leak some timing information, in particular, due to their inability to fully control how microarchitectural state affects program timing. {\em To demonstrate that even small amounts of timing leakage creates significant vulnerabilities, we show that a state-of-the-art timing leakage mitigation technique is still susceptible to the decade-old Bernstein cache attack ~\cite{Bernstein}.} We also show in this section that the Ozone execution environment, detailed in the following section, is not vulnerable to Bernstein's cache attack (or any other timing-based attack).

The Bernstein cache attack works by repeatedly running an AES encryption kernel while only changing a single byte of input plain text and then inferring key information based on how that single change affects cache performance. As such, there is no external adversary to build protections against, instead, the AES kernel's cache experiences interference from itself (in the form of capacity and conflict misses), which results in timing variations that expose key information.
\vspace{-1em}
\begin{figure}[H]
%[caption=Equations for First Round AES Encryption,frame=tlrb,mathescape=true]
\caption{Equations for First Round AES Encryption.\ignore{: \textnormal{In this equation, \(K\) is the encryption key and \(n\) is an input data to be encrypted. The equations use four tables (\(T_0 - T_3\)). The indexes of the tables are functions of the key bytes, \(K[i]\).}}}
\vspace{-4mm}
\normalsize
{\[Legend: K = 16~byte~key,~n = 16~byte~plain~text\]
\[s[i] = K[i] \oplus n[i], where~i = 0~to~15\]
\[ t_0 = T_0[s[0]] \oplus T_1[s[5]] \oplus T_2[s[10]] \oplus T_3[s[15]] \oplus x_0 \]
\[ t_1 = T_0[s[4]] \oplus T_1[s[9]] \oplus T_2[s[14]] \oplus T_3[s[3]] \oplus x_1 \]
\[ t_2 = T_0[s[8]] \oplus T_1[s[13]] \oplus T_2[s[2]] \oplus T_3[s[7]] \oplus x_2 \]
\[ t_3 = T_0[s[12]] \oplus T_1[s[1]] \oplus T_2[s[6]] \oplus T_3[s[11]] \oplus x_3 \]}
%t1 = 
\vspace{-2em}
\label{aes_equation}
\end{figure}

Specifically, the attack carefully manipulates indexing into the AES key tables. Figure \ref{aes_equation} shows the equations for the first round of AES encryption as implemented in the OpenSSL cryptographic library~\cite{openssl}. The implementation uses four key tables (\(T_0 - T_3\)). Accesses to these tables are derived from the AES key. In the equations, (\(K\)) is the encryption key and (\(n\)) is an input data to be encrypted. The table indexes are created by XOR'ing a key byte (\(K[i]\)) with a plain text byte (\(n[i]\)).

To demonstrate how even attenuated timing leakage leaves programs vulnerable, we perpetrated Bernstein's cache attack on an AES kernel with state-of-the-art protections while running on a simulated processor model. The AES program's protections and cache configuration are similar to the one's used in~\cite{Braun}. The program is protected such that it contains no input-dependent control, including a fixed number of loop iterations. In addition, the AES kernel starts execution with a flushed cache and reset branch predictor state.  The simulated caches are a 32KB L1 data cache, a 32KB instruction cache and a 256KB L2 shared cache. We used the Gem5 microarchitectural simulator for our experiments ~\cite{Gem5}. The AES encryption process is the only running process and the time measurements are total execution times in cycles.

%\ignore{\tma{The baseline is \cite{Braun}, correct?  Need to make that clear here, and need to make it clear that their technique is a very advanced one that accounts for timing leakage through control and data... and it still leaks valuable information because they cannot fully control timing variation in the microarchitecture.}}

\begin{figure}[t]
\includegraphics[trim={2cm 5cm 2cm 5cm},width=0.45\textwidth]{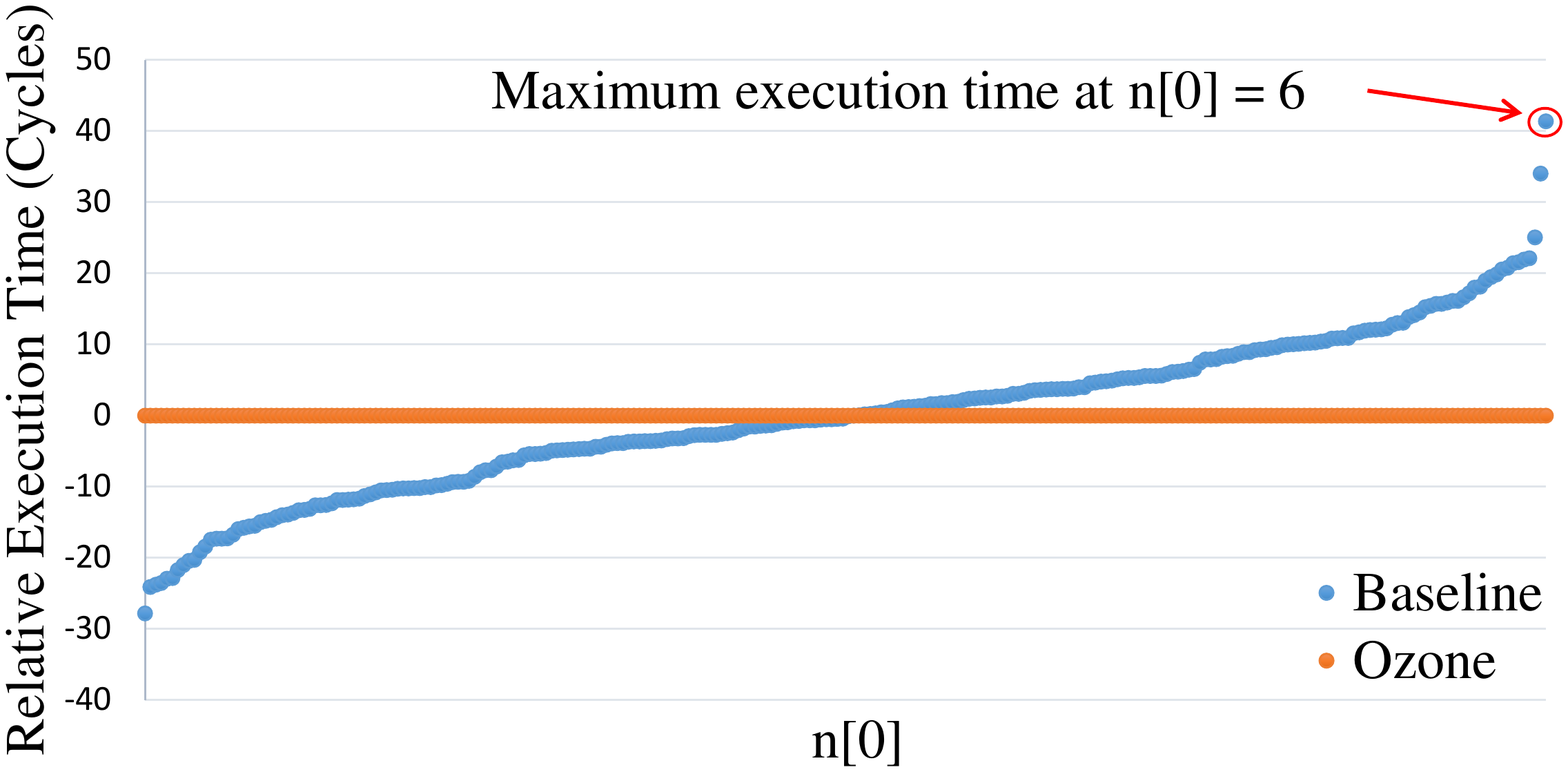}
\centering
 \caption{Relative Execution Time of AES Encryption: \textnormal{The graph shows the sorted relative total execution times for AES encryption as the value of an input data byte, \(n[0]\), varies from 0 to 255, for a state-of-the-art protection mechanism~\cite{Braun} and Ozone.\ignore{The baseline protects sensitive code by allocating a separate core to the code. The time variation for the baseline is a result of different cache conflict patterns that depend on the value of \(n[0]\). The value of \(n[0]\) with the maximum execution time (6 in this case) is used as a reference to recover the unknown key byte, \(k[0]\).}}}
\hrulefill
\label{attack_time}
\end{figure}

\ignore{\begin{figure}[t]
\includegraphics[width=0.4\textwidth]{actual_predicted.pdf}
\centering
\caption{Relation between actual and predicted values for \(K[0]\): \textnormal{The graph shows the relation between actual and predicted values for selected key byte values (\(K[0]\)). The values are predicted using the maximum execution value obtained in Figure~\ref{attack_time} as a reference.}}
\hrulefill
\label{actual_predicted}
\end{figure}}

Figure \ref{attack_time} shows relative total execution time for AES encryption as a function of \(s[0]\) $=$ (\(K[0] \oplus n[0]\)). Each point in the graph is obtained by averaging the total execution time for \(2^{10}\) 128-bit keys with a fixed value of \(n[0]\). The total execution times are given relative to the average execution time for all values of \(n[0]\). On the graph, we can see that there is a slight variation in execution time across different values of \(n[0]\). {\em This timing variation was sufficient to allow us to fully recover the AES kernel's secret key.} For example, from the graph, the maximum execution time is observed when \(n[0]\) is 6. Therefore we can conclude that, for any unknown key byte \(K[0]\), the maximum execution time is observed at \(K[0] \oplus n[0]\) = 6. By choosing different values for \(n[0]\) and measuring total execution time, we could infer the value of \(K[0]\). We then repeated this process on the remainding key bytes to fully recover the AES secret key.

Figure \ref{attack_time} also shows that the Ozone execution, running with the same inputs as the baseline experiment, executes precisely the same number of cycles regardless of changes to the input. This property hold for both changes in the plain text and the secret key values. Consequently, there is zero timing leakage for the Ozone executions, and it is not possible to implement Bernstein's attack on an Ozone execution.

\ignore{Figure~\ref{actual_predicted} shows the relation between the actual and predicted values of \(K[0]\) using the maximum execution time value in Figure~\ref{attack_time} as a reference. As can be seen from the graph, most of the key byte values are predicted correctly the first time. And, then the prediction is wrong, it is typically off by a small delta. The accuracy of the prediction can be improved by selecting more than one value as a reference (for example using the highest {\em n} execution time values). Furthermore, similar technique can be applied to the rest of the key bytes to retrieve the entire key. \tma{Interesting result, but I don't think it is needed to make our point, I think Figure 3 and this text should be removed, or make a more specific point that helps the reader later in the paper. My concern is that it is non-intuitive that XOR operations would not result in widly different predictions when they were wrong.}}

%\ignore{\tma{Make it very clear why timing variation still exists... it is due to microarchitectural effects that they cannot control, correct?}}

To summarize, in this section we showed that current mitigation techniques can only limit timing side channels due to the inherent variable latencies associated with microarchitectural features such as caches. We also showed that even highly attenuated timing leakage can still reveal sensitive information. As such, to fully stem the vulnerabilities of timing-based side-channel attacks, an architecture that eliminates all sources of timing variation is required.
%\vspace{-1mm}
\section{The Ozone Architecture}
%\vspace{-1mm}
The goal of Ozone is to eliminate all sources of timing side channels and achieve zero timing leakage. This is achieved by insuring that the execution of Ozone code ({\em i.e.}, the portion of code that we want protected from timing-based side-channel attacks) executes the same number of cycles regardless of its inputs. Moreover, the code's run time cannot change due to the activities of other threads and processes. By doing this, an attacker will not be able to gain any input-related information about code executing inside the Ozone execution resource. In this section we will describe how the Ozone architecture achieves zero timing leakage.

\begin{figure}[t]
\includegraphics[trim={0 0.5cm 0 0.5cm},scale=0.5]{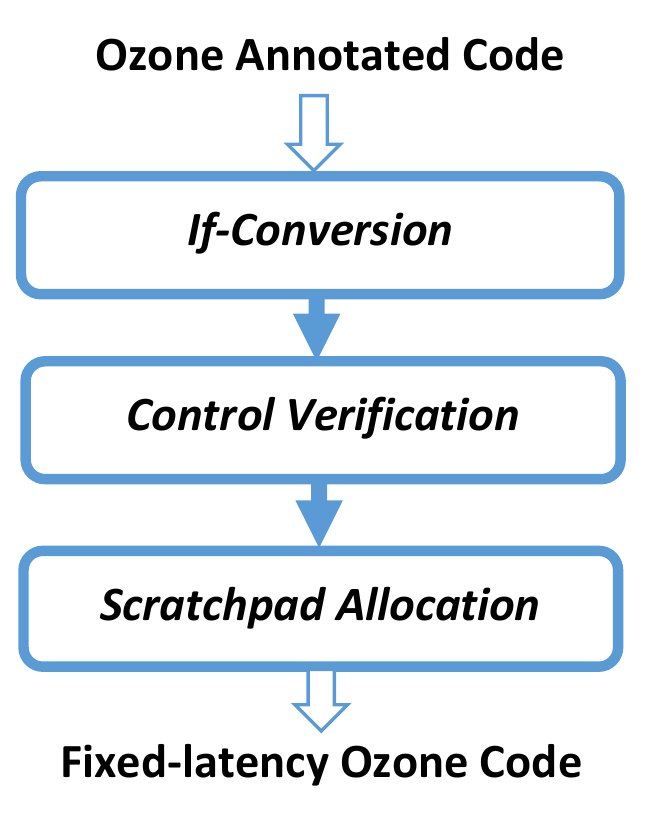}
\centering
\caption{The Ozone Compiler: \textnormal{The Ozone compiler, built as a back-end phase of LLVM, ensures that the code marked for Ozone execution {\em i}) has input-independent control flow via if-conversion and control verification, and {\em ii}) only accesses scratchpad memory. \ignore{With limited additional microarchitectural support, the Ozone compiler code will execute efficiently with fixed latency for all inputs.}}}
\hrulefill
\label{ozone_compiler}
\end{figure}

\ignore{There are two challenges that need to be addressed to achieve zero timing leakage. First, the code needs to be written (or compiled) to execute a fixed number of instructions on a fixed code path regardless of its inputs. This aspect of the code is enforced by the Ozone compiler, shown in Figure~\ref{ozone_compiler}. Second, when the program is invoked, it must start from a known fixed initial microarchitectural state, instructions must not exhibit variable latency, and the Ozone code's execution must be completely isolated from other threads and processes. Finally, we want to achieve this with minimal cost and performance impact and in a way that can be easily integrated into existing microarchitecture. In this section we will describe how we address each of these challenges in the Ozone architecture.}

\subsection{Threat Model}
In this work, we model the adversary as an unprivileged process that uses time variation in program execution to extract sensitive information. We assume the adversary has access to fine-grained timing information and microarchitectural statistics such as CPU cycle count and cache miss rates through performance monitoring counters.  We assume the hardware, hypervisor and operating system are trusted so that the adversary doesn't have direct read-write access to memory that is not allocated to it. The system compiler is not trusted, instead a small trusted verifier is used to verify Ozone code generated by the compiler. Finally, we assume the attacker will only pursue timing-based side-channel attacks, thus we consider other side channels, such as power and electromagnetics, beyond the scope of this work.

\subsection{Removing Input-Related Code Variability}
The first step in achieving zero leakage is eliminating all sources of variable latency within the Ozone code. Assuming fixed-latency instructions, this can be achieved by dynamically executing the same trace of instructions independent of input data. To guarantee the same dynamic trace, Ozone code control flow must be independent of its inputs, which means {\em i)} control hammocks ({\em e.g.}, IF statements) are not allowed, and {\em ii)} loops must execute a constant fixed number of iterations.

Control hammocks are removed by the Ozone compiler via if-conversion of CFG hammocks, shown in Figure \ref{ozone_compiler}. With if-conversion, all operations (\textit{except stores}) on either side of a conditional statement are executed regardless of the condition. Store operations need special treatment because they must avoid the potential side-effects of the false-predicate code. Stores are handled using a special \textit{conditional assignment function}, CMOV. \ignore{, similar to the IF-conversion implemented by~\cite{Rane}} Listings 1 to 4 show two examples of if-conversions that are common. Listing 1 shows a control hammock with stores in both control paths. In this example, prior to if-conversion, one of the two operations in the \textit{if} and \textit{else} bodies are evaluated based on the condition \textit{input==0}. After if-conversion (as shown in Listing 2), both operations are evaluated and a conditional assignment function is used to select the final value to store. Listing 3 shows an unbalanced control hammock (\textit{if-then construct}), where only one path performs a store operation. In this case, the conditional branch is eliminated by evaluating the statement inside the \textit{if} body and by conditionally storing the new result or the old value of \textit{y} based on the condition (as shown in Listing 4).\\

\noindent\begin{minipage}{.2\textwidth}
\small
\begin{lstlisting}[caption={If-Then-Else Conversion - Original Code},captionpos=b,frame=tlrb,mathescape=true]{Name}
if(input == 0){
  $y$ = $a^2$;
}
else{
  $y$ = $a^2$ * $b$;
}
\end{lstlisting}
\end{minipage}\hfill
\begin{minipage}{.25\textwidth}
\small
\begin{lstlisting}[caption={If-converted Code with Fixed Control},captionpos=b,frame=tlrb,mathescape=true]{Name}
$tv$ = $a^2$;
$fv$ = $a^2$ * $b$;
pred = (input == 0);
$y$ = CMOV(pred,$tv$,$fv$);
\end{lstlisting}
\end{minipage}

\noindent\begin{minipage}{.2\textwidth}
\small
\begin{lstlisting}[caption={If-Then Conversion - Original Code},captionpos=b,frame=tlrb,mathescape=true]{Name}
if(input == 0){
  $y$ = $a$ * $b$;
}
\end{lstlisting}
\end{minipage}\hfill
\begin{minipage}{.25\textwidth}
\small
\begin{lstlisting}[caption=If-converted Code with Fixed Control,captionpos=b,frame=tlrb,,mathescape=true]{Name}
$tv$ = $a$ * $b$;
pred = (input == 0);
$y$ = CMOV(pred,$tv$,$y$);
\end{lstlisting}
\end{minipage}

\begin{lstlisting}[caption={Conditional Store Function for x86},captionpos=b,frame=tlrb]{Name}
 CMOV (pred, tv, fv){
  mov    tv,%rax //Load true value
  cmp    $0,pred //Check if false
  cmove  fv,%rax //Load false value?
  return %rax }  //Return selected value
\end{lstlisting}

The conditional assignment function (CMOV) used in the if-conversions takes three inputs: a predicate value (\textit{pred}), a true value (\textit{tv}) and a false value (\textit{fv}), and returns either the true value or the false value based on the value of the predicate. It uses conditional move instructions such as the \textit{CMOVcc} instruction from the x86 architecture~\cite{Intel_sdm} to eliminate conditional branches. Listing 5 shows an example implementation of our conditional store for the x86 architecture.

After if-conversion by the Ozone compiler, all operations are performed regardless of their predicate conditions. In doing so, there can be cases where unchecked operations could produce exceptions such as division-by-zero. The Ozone compiler relies on the programmer to make sure that such faulty operations do not occur. Since stores in the Ozone execution environment always target the scratchpad (detailed below), these stores are not subject to page faults, thus making it trivial to ensure fault-free code. \ignore{If a more robust solution were desired for safe code execution, one could easily adopt one of the many solution from the VLIW compilation arena, {\em e.g.}, store pre-checks or page "tickles" (see \cite{Fisher} for examples).}

Figure~\ref{ozone_compiler} summarizes the Ozone compilation process. The input to the compiler is a programmer-annotated source code. Programmers annotate functions they want protected from timing-based side-channel attacks. In the first stage of the compilation process, conditional branches are eliminated by if-conversion. If-conversion is done on all annotated functions and functions called from annotated functions. Then, the control verification stage verifies that there are no conditional branch instructions in the Ozone code except for loop conditions with a fixed number of iterations and no early exits through \textit{break} and \textit{continue} statements. Finally, all verified Ozone code and associated data is placed in separate code and data sections that map to scratchpad memory. At the end of compilation, code with a fixed control path and data accesses to a scratchpad memory is produced; with support of the Ozone microarchitecture, this code will execute with a fixed latency regardless of its inputs.

%%%%Using if-conversion to eliminate conditional branches %%%%

%%(a)
%if(input == 0){
	%y = a * b;
%}
%else{
	%y = a^2 * b;
%}

%tmp0 = a * b;
%tmp1 = a^2 * b;
%cond = (input == 0);
%cmov(cond, tmp0, tmp1);

%%(b)
%if(input == 0){
	%y = a * b;
%}
%tmp0 = a * b;
%cond = (input == 0);
%y = cmov(cond, tmp0,y);

%%%%%%%%%%Conditional function

\begin{figure}[t]
\includegraphics[trim={0 0.5cm 0 0.5cm}, width=0.5\textwidth]{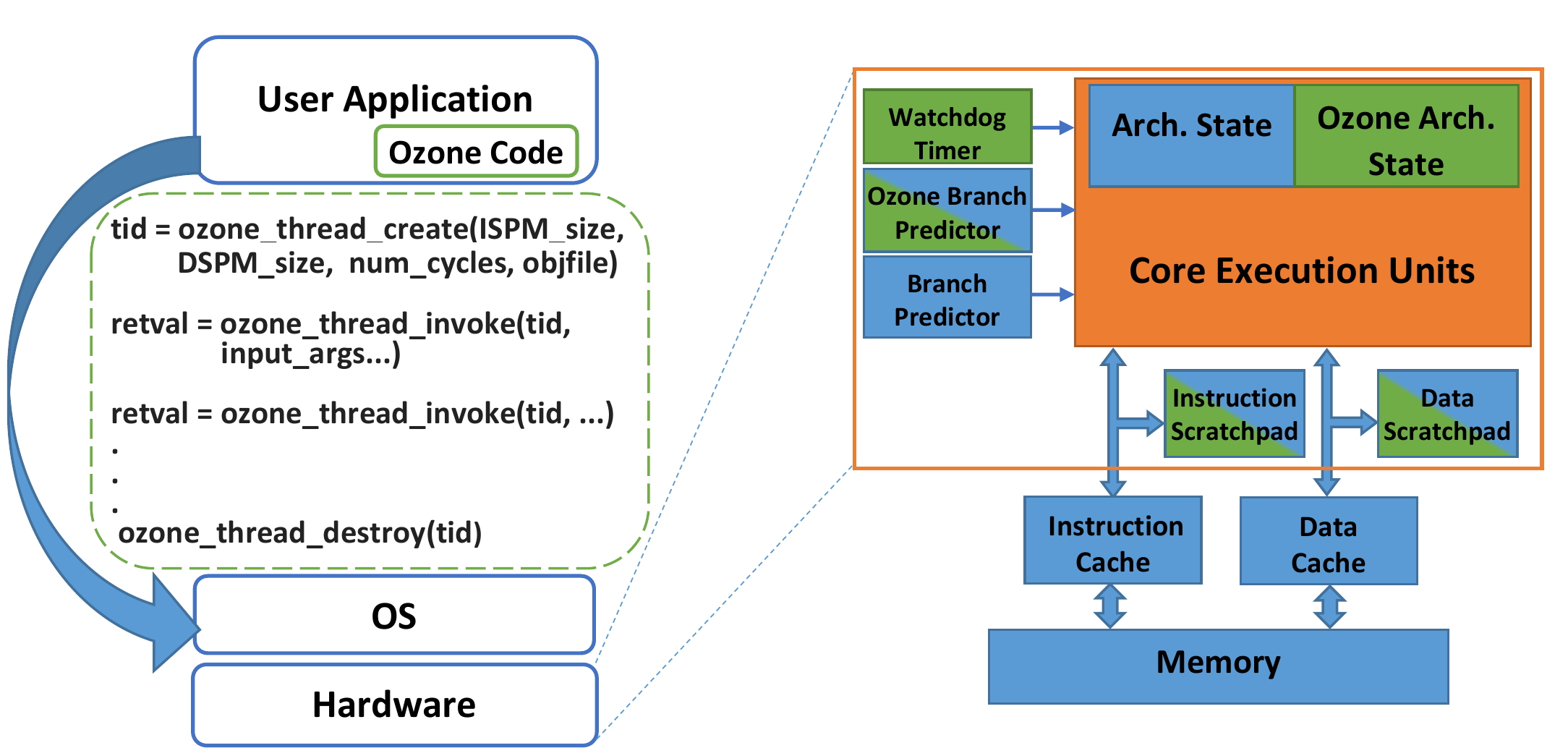}
\centering
\caption{The Ozone Architectural Extensions: \textnormal{The Ozone execution resource ensures zero timing leakage for execution on a modern microarchitecture using a few minor extensions to an existing core design. \ignore{To ensure fixed latency contention-free access to instruction and data memory, Ozone threads only utilize the instruction and data scratchpad memories (i.e., ISPM and DSPM). Microarchitectural resource contention attacks are prevented by giving Ozone threads exclusive access to the core's execution units and by utilizing the Ozone branch predictor, which is a simple stateless always-taken predictor.} }}
\hrulefill
\label{ozone-arch}
%\vspace{-6mm}
\end{figure}

\subsection{Eliminating Timing Leakage with Ozone}
Even after eliminating input-related control flow in Ozone code, the code will still be vulnerable to timing-based side-channel attacks on a traditional microarchitecture due to contention caused by resource  sharing.\ignore{ For instance, an aggressor could greedily snatch up resources and use {\em its own} performance as a gauge of what the victim thread is executing. Consequently, zero timing leakage requires additional support from the microarchitecture.} \ignore{In this section, we discuss how the Ozone architecture eliminates the variable latency due to the microarchitecture.}

\ignore{On modern processors, resources such as execution units, instruction and data caches, and branch predictors are shared among threads and processes. Resource sharing creates contention which can be used by an aggressor to infer the activities of another thread's execution.} Ozone eliminates contention by {\em i)} gaining exclusive access to the core's execution units during execution of Ozone code, {\em ii)} using fixed-latency instruction and data scratchpad memories (ISPM and DSPM) instead of caches and {\em iii)} ensuring all Ozone threads start execution with a known fixed microarchitectural state.

Figure~\ref{ozone-arch} shows the microarchitecture of a processor with an Ozone hardware thread. The processor components are divided into three groups. The first group of resources is exclusively used by the Ozone thread (shown in green in Figure~\ref{ozone-arch}) including the Ozone architectural registers and the watch-dog timer (WDT). The second group, shown as half-green and half-blue, is used by the Ozone execution resource exclusively for the lifetime of an Ozone thread. These resources include the instruction and data scratch-pad memories (ISPM and DSPM) and the Ozone branch predictor. Once all Ozone threads are destroyed, these resources can (optionally) be used by the rest of the system. For instance, the ISPM and DSPM could be allocated from a way of existing caches, for use by an Ozone thread. The final group of resources, shown in blue in the figure, are explicitly not used by Ozone threads as contention on these resources could enable timing attacks. These off-limit resources include the main branch predictor, caches, other thread states, and DRAM.

An application that wants to use the Ozone resource, first creates an Ozone thread by specifying the resources it requires ({\em ozone\_thread\_create}). The resources requested include instruction and data scratchpad memory sizes (including maximum stack space required for Ozone code execution), the fixed number of cycles required to execute the Ozone code, and a pointer to the Ozone code to load into the scratchpad memory. Following the creation of a new Ozone thread, the OS allocates instruction and data scratchpad memory space (including stack space in the data scratchpad), zeros out the allocated scratchpad memory, copies all read-only data into the data scratchpad and the Ozone code to the instruction scratchpad. The OS then returns a handle to the Ozone thread, which is used by the application to invoke the thread. 

%Figure XXX
\ignore{A call to an Ozone code function invokes a system call with the following parameters:
\begin{itemize}
\item The start and end addresses of the code section for the Ozone code function
\item The start and end addresses of the data section for the Ozone code function
\item The total execution cycles for the Ozone code function
\item The maximum stack space required
\item The input arguments to the Ozone code function
\item Pointer to memory location to put the results of the Ozone code function execution
\end{itemize}}

When the Ozone thread is invoked by the main program (via {\em ozone\_thread\_invoke}), the processor switches to the Ozone thread context. This thread switch forces a flush of the processor pipeline (to eliminate reservation station contention with the previous thread), and then enables the Ozone branch predictor with a fixed initial state. If the Ozone branch predictor is a static predictor ({\em e.g.}, predict always-taken), this predictor initialization step can be skipped. The hardware context is then switched to the Ozone thread and execution begins. A watchdog timer (WDT) in the processor is started that keeps track of the expected number of execution cycles for the Ozone thread invocation. When the WDT expires, the timer interrupts the Ozone thread, stopping execution. If execution completed in the same cycle the WDT expires, the result of the execution is returned, otherwise, the Ozone thread is terminated as it is running too long or not long enough.

\section{Experimental Evaluation}
% * <austin@umich.edu> 2016-03-30T02:07:26.747Z:
%
% ^.
\ignore{In this section we evaluate the security and performance of the Ozone architecture. We start by detailing the implementation of the hardware and software components of Ozone and the benchmark applications used for our evaluation.} 
\subsection{Ozone Implementation}
The Ozone compiler is composed of if-converter and control verifier stages. The compiler stages are implemented as IR passes on the x86-targeted LLVM compiler~\cite{LLVM}. The if-conversion is implemented by replacing stores by a call to the conditional assignment function as detailed in Section 3. \ignore{After if-conversion, the control verifier makes sure that there are no conditional branches in the entire Ozone code except for fixed iteration count loops. This check is made for the root Ozone function, plus any function called by it directly or indirectly. Scratch pad memory assignment is implemented using special code and data sections, which are then mapped to a specific address using a modified linker map.}
%Prior to automatic if-conversion, the kernels are hand optimized.

The Ozone microarchitecture is modeled using the Gem5 microarchitectural simulator~\cite{Gem5}. Table \ref{config-table} lists the configuration for the baseline \ignore{in-order and} out-of-order core. Ozone is integrated into the simulator as a special thread context, with the state detailed in Section 3. Additionally, an always-taken predictor is implemented for stateless prediction of Ozone code branches, and instruction and data scratchpads are integrated into the simulator at a fixed address.
\bgroup
\def\arraystretch{1.5}
\begin{table}[]
\centering
\caption{Baseline Microarchitecture Configuration.\ignore{: \textnormal{This table lists the baseline configuration used for the Gem5-based simulations used to perform security and performance analysis.}}}
\label{config-table}
\begin{tabular}{| l | l |}
\hline
L1 Instruction Cache Size & 32kB  \\ \hline
L1 Data Cache Size        & 32kB  \\ \hline
L2 Unified Cache Size    & 256kB \\ \hline
Cache Associativity    & 4-way \\ \hline
Issue Type     & Out-of-order \\ \hline
Issue Width     & 8 \\ \hline
Branch Predictor Type     & Tournament \\ \hline
Branch Predictor Size     & 56kB \\ \hline
\end{tabular}
\end{table}
\egroup

\subsection{Benchmark Applications}
%Table showing memory requirements for each benchmark 
The benchmark applications used in our evaluation constitute security sensitive applications that have previously been subject to side-channel attacks. They are adapted directly from widely used libraries: OpenSSL~\cite{openssl}, GDK~\cite{gdk} and glibc~\cite{glibc}. {\em Our intent with this benchmark suite is to demonstrate the broad utility of the Ozone execution environment, thus, we have ported into it all of the codes we could find that have been previously attacked with timing-based side channels.} The benchmark applications include AES encryption, RSA decryption, SHA512 hash kernel, and a key-mapping function (GDK-Keymap). Table \ref{code-details} lists references to timing side-channel attacks for each of the benchmark applications.

\begin{table}[]
\centering
\caption{Analyzed Benchmarks with Scratchpad Memory Requirements. \textnormal{The table also lists references to attacks on each benchmark application.}\ignore{: \textnormal{This table lists the codes ported to the Ozone execution resource. The codes were selected because they have been the focus of side-channel based attacks in the past. Also listed are the instruction and data scratchpad memory requirements.}}}
\label{code-details}
\begin{tabular}{llll}
\toprule
\multicolumn{1}{c}{\textbf{Benchmark}} & \begin{tabular}[c]{@{}c@{}}\textbf{Instruction} \\ \textbf{Memory Size}\end{tabular} & \begin{tabular}[c]{@{}c@{}}\textbf{Data}\\ \textbf{Memory Size}\end{tabular} & \begin{tabular}[c]{@{}l@{}}\textbf{Attack}\\ \textbf{Examples}\end{tabular} \\ \midrule

AES-CBC &             6.627K      &                  12.16K	   &  \cite{Bernstein,Bonneau,Tromer} \\
AES-XTS                                &      6.382K                                                              &    12.16K                      &  \cite{Bernstein,Bonneau,Tromer}        \\
GDK-Keymap                             &             1.692K                                                       &       6.728K   &            \cite{Gruss}                                     \\
RSA                                    &        13.909K                                                            &         13.311K             &         \cite{Kocher,Yarom,Yarom2}                                  \\
SHA512                                 &      3.308K                                                              &    4.288K                     &  \cite{Lawson}                                 \\ \bottomrule
\end{tabular}
\end{table}

Table \ref{code-details} shows the instruction and data memory requirements for the benchmark applications, including the maximum stack space required. As can be observed from the table, the kernels and their data can easily fit into small scratchpad memories. For these codes, a 32kB instruction scratchpad and a 64kB data scratchpad will hold all of these algorithms simultaneously (and eliminate the need for context switch saves), and a 16kB instruction and data scratchpad will comfortably hold any single application component.

\subsection{Performance and Security Evaluation}

\begin{figure*}[!t]
\includegraphics[width=0.9\textwidth]{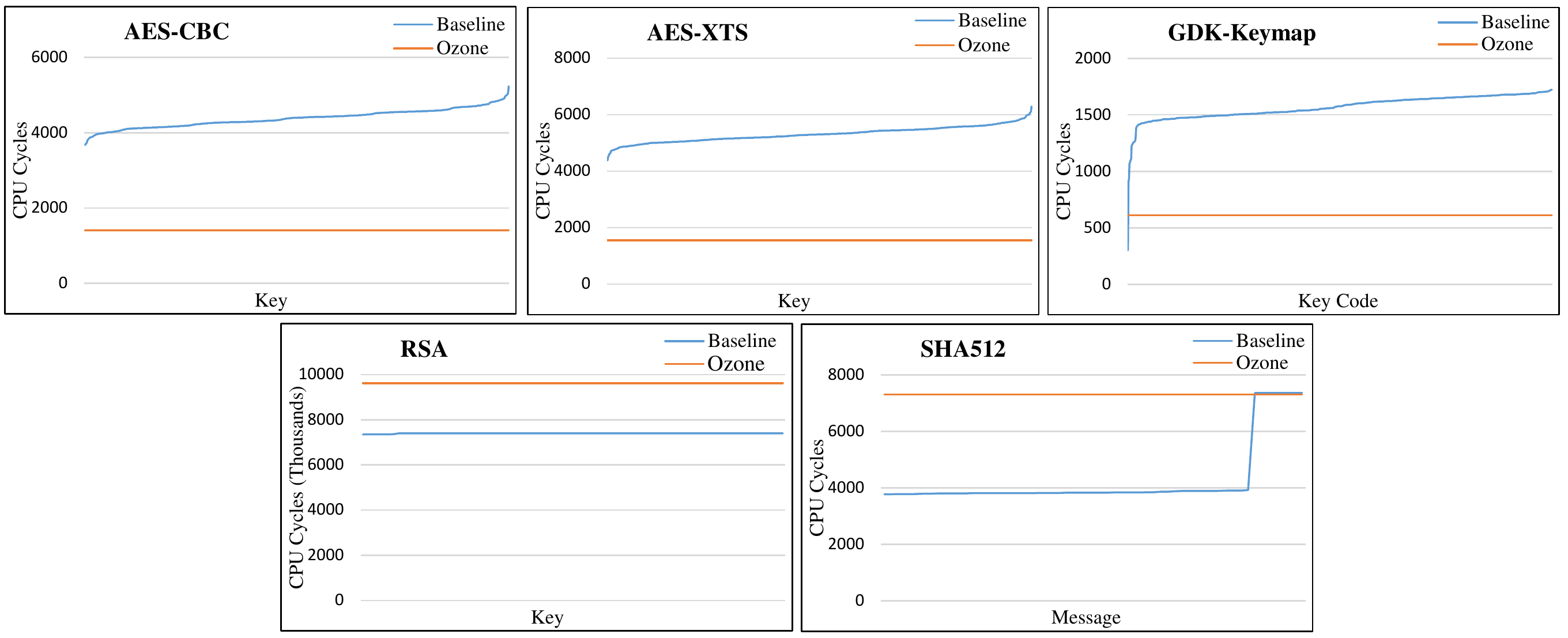}
\centering
\caption{Security Analysis of Ozone Executions: \textnormal{Each graph shows for each of the five benchmarks the performance variability across a wide range of inputs. The baseline architecture performance (in blue) is shown for varied inputs sorted from fastest to slowest. The Ozone lines are always horizontal, meaning that there is zero timing leakage as the timing is fixed for all tested inputs}}
\hrulefill
\label{cpu_cycles}
\vspace{-1em}
\end{figure*}

\ignore{To evaluate the security of the Ozone execution resource, it is necessary to demonstrate that it does not exhibit timing leakage of any form, for a wide range of execution scenarios.} We evaluated the security and performance of our benchmark codes running on the baseline out-of-order microarchitecture and in the Ozone execution resource across many random inputs. The inputs were varied in the following manner: AES-CBC encrypts with 1024 random keys with random 128-bit inputs, AES-XTC encrypts with 1024 random keys with random 256-bit inputs, GDK-keymap maps all possible 784 keyboard codes, RSA decrypts a random 128-byte message with 1024 random 1024-bit keys, and SHA512 hashes random inputs from length 1 to 128 bytes.

Figure \ref{cpu_cycles} shows the results of the security analysis of the Ozone execution resource. \ignore{Each graph shows the result for an individual benchmark.} The graphs show the performance of the baseline microarchitecture (in blue) and that of the Ozone execution (in orange) across a wide range of inputs, with the results sorted and plotted from fastest execution to the slowest.\ignore{ For the Ozone execution (shown in orange), all executions take the same number of cycles, thus, the results form a horizontal line on each graph.} For each of the baseline experiments, the branch predictor and caches are flushed prior to the start of execution.

It is interesting to note the timing leakage characteristics of the baseline out-of-order executions. In all cases, there is evidence of the performance of the program changing as the inputs are varied. As expected, however, the Ozone executions exhibit zero timing leakage, by demonstrating a fixed execution latency across all inputs. \ignore{This result stems from the fixed control path taken by the Ozone codes, leading to a fixed sequence of instructions regardless of how the inputs vary.} The Ozone executions are also immune to aggressor thread perturbations since they utilize static always-taken branch speculation and avoid use of the cache, all the while maintaining exclusive access to hardware resources for the duration of the execution.

For most of the codes, exclusive access to the resources (which also disables interrupts for the core) is a short duration, on the order of 500-7500 cycles. The one outlier is RSA, which requires more than 9.5M cycles. This would likely be too long to completely ignore interrupts and exceptions. Thus, a system integrating Ozone execution would likely dedicate it to only a subset of cores, leaving the remaining cores in the system to provide timely response to interrupts and exceptions.

The performance of the Ozone execution compares favorably to the baseline out-of-order executions. In more than half of the cases (3 of 5), the Ozone execution was more efficient than the baseline execution. This is possible due to faster scratchpad accesses ({\em i.e.}, no cache misses), more responsive branch prediction ({\em i.e.}, the baseline tournament predictor takes much longer to warm up compared to Ozone's stateless always-taken predictor), and fewer difficult-to-predict if-statements (which are removed during if-conversion). Conversely, the executions of {\em RSA} and {\em SHA512} were less efficient. In the case of {\em RSA}, this was primarily due to significant overheads incurred during if-conversion, which requires that both the true and false code of an IF statement be executed for all occurrences. In the case of {\em SHA512}, the Ozone implementation pads out the password being hashed to always 128 bytes, thus, it performs the worst-case length hash for all inputs. This extra work is unavoidable if one wants to conceal password length.

\subsection{Ozone Resource Cost Assessment}
Table \ref{cost_table} estimates the bit-area costs of implementing the Ozone execution resource. The table shows the various components necessary to implement Ozone threads and their cost in bits. The table also indicates if the component is required, as some of the components, in particular the scratchpad memories, can be temporarily borrowed from existing microarchitectural resources (namely, the cache).

The overall cost of implementing the Ozone execution resource can be kept quite low. In particular, if the instruction and data scratchpad memories are acquired via reconfiguration of the instruction and data caches ({\em e.g.}, claim one way of these caches), then the overall cost for a single Ozone thread context is only 11 bytes (plus baseline thread state). With additional Ozone thread contexts and dedicated instruction and data scratchpad memories, implementation costs will grow accordingly. However, even an Ozone implementation with full dedicated resources will only require silicon area on the order of the L1 instruction and data caches of an individual core (approximately 96kb of storage).

\begin{table}[]
\centering
\caption{Cost Assessment. \textnormal{This table estimates the bit-area cost of the Ozone execution resource. \ignore{Included components are the Ozone execution mode bit, the additional thread context state, and instruction and data scratchpad memories. Note that the scratchpad memories can be made optional by reassigning one way of the instruction and data caches for use as dedicated Ozone RAM arrays.}}}

\label{cost_table}
\begin{tabular}{@{}lcc@{}}
\toprule
\multicolumn{1}{c}{\textbf{Addition}} & \textbf{Cost (in bits)} & \textbf{Optional?} \\ \midrule
Ozone mode bit               & 1              & No        \\
Ozone thread context         & 80             & No        \\
Instruction Scratchpad       & 32k x 8        & Yes       \\
Data Scratchpad              & 64k x 8        & Yes       \\
\toprule
\multicolumn{3}{l}{Expected Total Cost: 11 bytes ...to... 96kB}  \\ \bottomrule
\end{tabular}
\end{table}

\section{Conclusion}
\ignore{Timing-based side channels have proven to be a very effective approach to extracting secrets by simply observing the runtime of security sensitive executions. Attacks have demonstrated that caches, branch predictors, complex control, code sharing, among other things all lend themselves to efficient and effective side channel attacks. It's not surprising that today's computing devices are so susceptible to side channel attacks, when one realizes that {\em i)} optimizing the common case, {\em ii)} resource sharing, and {\em iii)} fine-grained performance analysis capabilities are the root causes of timing-based side channels.}

In conclusion, this work pushes the state-of-the-art in timing-based side-channel resistant execution forward with the Ozone execution resource. Unlike previous work, Ozone implements a {\em zero timing leakage} execution capability, and it does so with low area cost (as few as 11 bytes of state) and significantly less performance impact than previous non-zero timing-leakage proposals (no more than a 30\% performance loss for a modern microarchitecture). Our approach is to map a carefully prepared input-independent code sequence, built with the Ozone compiler, to a microarchitecture with stateless Ozone branch predictors and small instruction and data scratchpad memories. In this effort, we examine five benchmarks pulled from the literature as being particularly susceptible to side-channel attacks. The Ozone design allows these codes to efficiently execute with a fixed number of cycles, regardless of input, on even a complex microarchitecture. By eliminating all sources of timing variation in the Ozone execution environment, codes ported to it can rest assured that their executions do not leak secrets via timing channels.

Looking ahead, we see a great opportunity to address additional side channels with the Ozone execution resource. In particular, if the Ozone execution resource were moved into a physical co-processor, it could be designed with circuits to minimize power side channels, while still retaining its capability to execute code with zero timing leakage.

\section{Acknowledgments}
This work was supported in part by C-FAR, one of the six STARnet Centers, sponsored by MARCO and DARPA.



%\bibliographystyle{IEEEtran}
\vspace{6pt}
\scriptsize{
\bibliographystyle{plain}
\bibliography{references}
}

\end{document}